\documentclass[12pt,english]{article}

\renewcommand{\Im}{{\rm Im \, }}
\renewcommand{\Re}{{\rm Re \, }}

\newcommand{\AdS}{{\rm AdS }}

\usepackage[utf8]{inputenc}
\usepackage{geometry}
\usepackage{youngtab}

\usepackage{babel}

\usepackage{verbatim}
\usepackage{prettyref}
\usepackage{amsmath}
\usepackage{amssymb}
\usepackage{mathabx}
\usepackage{mathrsfs}
\usepackage{graphicx}
\usepackage{esint}
\usepackage{physics}
\usepackage{slashed}
\usepackage{subcaption}
\usepackage{array}
\newcolumntype{C}{>{$}c<{$}}
\usepackage{dsfont}
\usepackage{xcolor}
\usepackage{color}
\definecolor{darkgreen}{rgb}{0,0.5,0}
\definecolor{darkblue}{rgb}{0,0,0.6}
\definecolor{purple2}{rgb}{0.4,.2,0.7}

\newcommand{\newrev}[1]{{#1}}
\usepackage[colorlinks=true,citecolor=purple,linkcolor=purple2,urlcolor=purple2]{hyperref}
 	
\numberwithin{equation}{section}
\numberwithin{figure}{section}
\numberwithin{table}{section}

\def\IR{{\mathbb R}}

\newcommand{\rme}{\mathrm e}
\newcommand{\rmd}{\mathrm d}
\newcommand{\rmi}{\mathrm i}

\addtocounter{page}{-1}

\begin{document}
\vspace*{2cm}
\begin{center}
{\LARGE \bf  D(\texorpdfstring{$-$1}{-1})/D7 bound states in supergravity}

    \vspace{1.5cm}
    {\large Gongrui Cheng$^{1}$, Klaas Parmentier$^2$, Thomas Van Riet$^1$}\\
    \vspace{1cm}

    {$^1$\emph{
        Instituut voor Theoretische Fysica, K.U. Leuven,\\
        Celestijnenlaan 200D B-3001 Leuven, Belgium\\
    }
    $^2$\emph{Department of Applied Mathematics and Theoretical Physics,\\
University of Cambridge, Cambridge CB3 0WA, UK\\}
    }
    
    \vspace{2cm}
    
    {\bf Abstract}\\
    \end{center}
    \newrev{We construct D($-$1)/D7 bound states in Euclidean type IIB supergravity motivated by the string-theory setup of Bill\`o et al., whose open-string description realizes instantons in eight-dimensional super Yang--Mills theory. Previous supergravity constructions either left unresolved issues concerning ten-dimensional Euclidean supersymmetry or relied on partial smearing to codimension one. We classify the local codimension-two solutions by showing that the supersymmetric field equations reduce to two holomorphic functions subject to a single complex constraint. The candidate bound-state solutions require complementary D3-brane fluxes, consistent with the D7 worldvolume analysis. A particularly interesting solution has constant dilaton and connects the vacuum at infinity to a controlled scaling limit in which the geometry becomes $\mathbb R\times S^1\times\mathbb T^8$. 
    }

{\small }

\setcounter{tocdepth}{2}
\thispagestyle{empty}
\newpage


\section{Introduction}
Despite the long history of constructing brane-type solutions in supergravity, a description of a D($-$1)/D7 bound state was lacking until recent attempts \cite{Aguilar-Gutierrez:2022kvk, Reymond:2024mwe}, which were inspired by a string theory analysis of the D($-$1)/D7 system \cite{Billo:2021xzh}. One motivation for the study of such bound states arises from an interest in classifying instantons in 8D SYM, which unlike in 4D SYM (described by D($-$1)/D3), remains an unsolved problem \cite{Nekrasov:2017cih}. 

Another motivation is moduli stabilisation, where instantons of various kinds --- D($-$1) and Euclidean D3 --- are expected to contribute to an effective scalar potential that lifts classical moduli. Such scenarios in IIB usually have D7 branes in the background, needed for tadpole cancellation and to induce certain gauge groups \cite{Ibanez:2012zz}. 

A further interest are the intriguing subtleties arising in D-brane systems with eight mixed Dirichlet–Neumann directions instead of the familiar four. Such setups reveal non-trivial physics, usually predicting extra branes needed for consistency. For instance, the D0-D8 bound state has a fundamental string (denoted as F1 here) stretching in between, which was predicted from string and field theory analyses \cite{Hanany:1996ie, Danielsson:1997wq, Witten:2000mf} and then rediscovered in supergravity \cite{Janssen:1999sa,Massar:1999sb, Imamura:2001cr, Bergshoeff:2003sy}. Interestingly, the supergravity description of the D($-$1)/D7 system of \cite{Billo:2021xzh} is \emph{not} the timelike T-dual of such a D0-D8-F1 bound-state, but is more akin to a D0-D4-D8 bound state, which has not been constructed before.\footnote{We are grateful to Iosif Bena and Dimitrios Toulikas for discussions on this matter.} 

A final, more heuristic motivation stems from holography \cite{Aguilar-Gutierrez:2022kvk, Reymond:2024mwe}. 
Unlike the D3-brane case, generic extremal D$p$-brane solutions have a running dilaton.
However, by constructing bound states of D$p$ branes with their magnetically dual D(6$-p$) branes, the dilaton can approach a constant in the throat, which may provide a smooth AdS near-horizon. The D1-D5 bound state is a famous example of this with an $\AdS_3\times S^3\times \mathbb{T}^4$ near-horizon limit having an SCFT$_2$ dual. Extrapolating this, one expects the near-horizon of a D($-$1)/D7 bound state, with the D7s reduced over a $\mathbb{T}^8$, to be a ``conformal matrix theory" obtained from adding string-interactions between the D7 and D($-$1) stack. It was conjectured in \cite{Aguilar-Gutierrez:2022kvk} that such a conformal matrix theory is described by the instanton moduli of \cite{Billo:2021xzh}.\footnote{A possible definition of matrix RG flows, and hence fixed points, is given by the so-called Brezin-Zinn-Justin RG flows \cite{Brezin_ZJ_1992, Zinn-Justin:2014wva} and its connection with holography is described in \cite{Guan:2026sjw}.}

In the rest of the paper, we present our Ansatz and derive the minimal set of equations a supersymmetric solution must satisfy. We find two general classes of solutions: genuinely codimension-two configurations, as well as a locally codimension-one solution generalizing~\cite{Reymond:2024mwe}.

\section{The Ansatz}\label{sec:ansatz}
As reviewed for example in \cite{Aguilar-Gutierrez:2022kvk}, reality conditions in Euclidean (super)gravity are subtle, even leaving aside the possibility that a real section of the complex superalgebra may not exist \cite{Bergshoeff:2000qu}. The appropriate reality conditions are determined by the physical observable: Euclidean solutions are saddle points contributing to specified matrix elements, in other words, to path integrals with specified boundary conditions \cite{Burgess:1989da}. For the D($-$1) brane~\cite{Gibbons:1995vg}, or its axion-wormhole relatives, one may work with an imaginary axion $C_0$ and impose Neumann boundary conditions on the background and its fluctuations \cite{Loges:2022nuw, Hertog:2024nys}. Alternatively, one may use a real $C_8$ in the Hodge-dual formulation with a Dirichlet boundary condition. Both amount to fixing the D($-$1) charge. Describing D7 charge instead leads to the opposite choice of reality and boundary conditions.

In this work we focus on finding a supergravity description of a D($-$1)/D7 bound state, as described in a worldsheet and 8D SYM context by \cite{Billo:2021xzh}. This implies we need a complex $C_0$ (or $C_8$). Further insisting on a real metric forces the inclusion of D3 branes \cite{Aguilar-Gutierrez:2022kvk, Reymond:2024mwe}.  It is reassuring that the condition for canceling the imaginary part of the Euclidean energy-momentum tensor forces D3-brane-like fluxes of the kind sourced by worldvolume gauge fluxes on the D7 brane, as required for consistency of the open-string picture \cite{Billo:2021xzh}.

All of this motivates the following Ansatz, in 10d Einstein frame,\footnote{The Einstein frame metric $\rmd s^2_E$ and string frame metric $\rmd s^2$ are related via $\rmd s^2_E= \rme^{-\phi/2}\,\rmd s^2$.}
\begin{align}\label{eq:ansatz}
& \rmd s^2_E = \rme^{2a(x,y)}(\rmd x^2 + \rmd y^2)+ \rme^{2b(x,y)}\sum^4_{i=1}\rmd \theta^2_i+\rme^{2c(x,y)}\sum^8_{i=5}\rmd \theta^2_i\,,\\
&   F_1 = \rmd U(x,y) + \rmi\, \rmd V(x,y)\,, \qquad \phi=\phi(x,y)\,,\\
&  F_5 = (1-\rmi \star)\mathcal{F}\, , \quad \mathcal{F}= \rmd \theta_{1234} \wedge (\rmd S(x,y) + \rmi\, \rmd T(x,y))\,.
\end{align}
The Euclidean D7 stacks are in the directions $\theta_1\ldots\theta_8$.
Note that, in this Ansatz, all variables $a,b,c,\phi, U, V, S, T$ are real functions of $x$ and $y$. The partially smeared solutions of \cite{Reymond:2024mwe} are found by dropping the dependence on either $x$ or $y$ in the metric, dilaton and field strengths.

In what follows, we will consider the coordinates $\theta_i$ to parametrize circles, such that the Euclidean D7s wrap a $\mathbb{T}^8$. For the sake of finding local solutions to the equations of motion, these topological properties are irrelevant, but they matter for charges and regularity. As explained in the introduction, our goal here is to construct the $p=-1$ entry in the chain $\AdS_{p+2}\times S^{p+2}\times \mathbb{T}^{6-2p}$ of backgrounds obtained by taking a near-horizon of D$p$/D(6-$p$) bound states with the larger brane wrapped over the torus \cite{Aguilar-Gutierrez:2022kvk}.

\section{Supersymmetry and equations of motion}
In this section, we derive the BPS conditions for the Ansatz \eqref{eq:ansatz} and find a minimal set of equations that must be satisfied in order to claim a supersymmetric solution. The coupled second-order equations of motion for the eight functions $a,b,c,\phi,U,V,S,T$, all depending on $x$ and $y$, are derived in Appendix \ref{sec:EOM}.
\subsection{BPS equations}\label{sec:susy}
We will now analyse the first-order PDEs resulting from the supersymmetry variations of the dilatino and gravitino (written in string frame): 
\begin{align}
 &   \delta_\epsilon \lambda = \frac12(\partial_\mu \phi - \rmi \rme^{\phi}F_{1\mu})\Gamma^\mu \epsilon = 0\,,\\
&   \delta_\epsilon\psi_\mu = D_\mu \epsilon +\frac{\rmi}{8}\rme^\phi( \Gamma^\nu F_{1\nu}+\frac{1}{2\cdot 5!}\Gamma^{ijklm}F_{5ijklm})\Gamma_\mu \epsilon=0\,,
\end{align}
with $D_\mu \epsilon = \partial_\mu \epsilon+\frac14 \omega_\mu^{ab}\Gamma_{ab}\epsilon$, where $\omega_\mu^{ab}$ is the string-frame spin connection. 

For our Ansatz we impose the following projectors:\footnote{More precisely, one may take
$\Gamma_{1234}\epsilon=\Gamma_{5678}\epsilon=\eta\epsilon$ with
$\eta=\pm1$. We choose $\eta=+1$; the opposite choice amounts to
$F_5\to-F_5$ and reverses both D3 charges.}
\begin{align}
  &\text{D($-$1)/D7:}\qquad   \Gamma_{xy}\epsilon = \rmi \epsilon\,, \label{eq:XYproj}\\
  &\text{D3/D3: }\qquad 
   \Gamma_{1234}\epsilon=\Gamma_{5678}\epsilon\,.\label{eq:D3proj}
\end{align}
For a chiral IIB spinor $\Gamma_{xy12345678}\epsilon = \rmi \epsilon$, so the D3/D3 relation follows from the D($-$1)/D7 projector and chirality. We nevertheless display it as it makes the relative orientation of the D3 sectors manifest.\footnote{The sign here is correlated with the relative sign in $F_5 = (1-\rmi \star)\mathcal{F}$. Indeed, the forms can be constructed from the spinors. Moreover, this matching of the signs is required for the $\Gamma \mathcal{F}$ and $\Gamma\star\mathcal{F}$ terms to add up in $\delta_\epsilon \psi_\mu$ (rather than subtract and cancel out). This is what allows for supersymmetric D3 brane solutions.} The real and imaginary parts of the dilatino variation then become
\begin{equation}\label{eq:dilatino}
    \partial_x U-\partial_y(V-\rme^{-\phi})=\partial_y U + \partial_x (V-\rme^{-\phi})=0\,.
\end{equation}
These can be recognized as Cauchy-Riemann equations, imposing holomorphicity \cite{lander}:
\begin{equation}\label{eq:hol1}
    \bar{\partial} \tau_1=0\,,\quad \tau_1 = U + \rmi(V-\rme^{-\phi})\,, \quad \bar\partial = \tfrac12(\partial_x + \rmi \partial_y)\,.
\end{equation}
We now analyze the gravitino variation. First, the $\theta_1$- and $\theta_5$-components yield equations which reduce (after using the dilatino variation \eqref{eq:dilatino}) to
\begin{align}
  & \partial_x(b+c)=\partial_y(b+c)= 0\,,\label{eq:gravi3}\\
 &  \partial_x S - \partial_y(T-\rme^{4b})=\partial_y S + \partial_x(T-\rme^{4b})=0\,. \label{eq:gravi4}
\end{align}
Remarkably, the second condition again imposes holomorphicity 
\begin{equation}\label{eq:hol2}
    \bar{\partial}\tau_2 = 0\, , \quad \tau_2 = S + \rmi (T- \rme^{4b})\,.
\end{equation}
Next, the $x$- and $y$-components impose (after having simplified using the dilatino variation \eqref{eq:dilatino} and the previously found gravitino variations)
\begin{align}
  & \partial \epsilon + \left(\frac12\partial a+\frac18\partial \phi\right)\epsilon  =0 \iff \epsilon(x,y) = \rme^{-a/2-\phi/8}\epsilon_0\, , \label{eq:gravi1}\\
  & \partial(2a-2b+\phi)+\rme^\phi \partial V +  \rme^{-4b}\partial T=0\,. \label{eq:gravi2}
\end{align}
In manipulating the gravitino equation, we assumed $\epsilon = \rme^f \epsilon_0$ with $f\in \IR$. However, it is possible that the spinor acquires a phase. In this case, \eqref{eq:gravi1} and \eqref{eq:gravi2} become
\begin{align}
     &  \partial_i \Re(f)+ \frac12\partial_i a+\frac18\partial_i \phi  =  \epsilon_{ij}\partial_j\Im(f) \, , \label{eq:gravi1b}\\
 &      \partial_i(2a-2b+\phi)+\rme^\phi \partial_i V +  \rme^{-4b}\partial_i T=4 \epsilon_{ij} \partial_j \Im(f)\,.\label{eq:gravi2b}
\end{align}
Note that taking the curl of \eqref{eq:gravi1b} immediately implies $\Im(f)$ is harmonic. 

\subsection{Minimal set of equations for a supersymmetric solution}\label{sec:bps2}
The standard lore is that the BPS equations plus Bianchi imply the equations of motion. Nevertheless, in IIB Euclidean supergravity some subtleties can be expected \cite{Bergshoeff:2000qu}. To this end, we check in Appendix \ref{sec:implied} which equations of motion are implied by the BPS equations derived above, and which must be separately imposed.  We summarize those conditions here.

Supersymmetry imposes the following conditions:
\begin{align}
    &\bar{\partial}{\tau_1}= \bar{\partial}\tau_2 = 0\,,\label{eq:susy1}\\
    & \partial(2a-2b+\phi) + \rme^{\phi}\partial V + \rme^{-4b}\partial T = 0\,,\label{eq:susy2}\\
    & b+c = \mathrm{ct}. \label{eq:susy3}
\end{align}    
with $ \tau_1 = U+\rmi(V-\rme^{-\phi})$ and $\tau_2 = S + \rmi (T-\rme^{4b})$ as defined in \eqref{eq:hol1} and \eqref{eq:hol2}. The constant in \eqref{eq:susy3} can be absorbed into the coordinate radii of the two tori, so we set $c=-b$ below.

If the spinor carries a phase, \eqref{eq:susy2} is replaced by \eqref{eq:gravi2b}. The real part of the Killing-spinor exponent $f$ is always determined by \eqref{eq:gravi1b}.\footnote{Integrability of \eqref{eq:gravi1b} implies directly that $\Im(f)$ is harmonic. Combining this with the integrability of \eqref{eq:gravi2b} then shows that the two orthogonality conditions displayed in \eqref{eq:eomxxa} are not independent: once either one is imposed, the other follows. See Appendix \ref{sec:implied}.}

In addition, in order to satisfy the equations of motion, we need
\begin{align}
&    \partial_i(\rme^{2\phi}\partial_i V) =\partial_i(\rme^{-8b}\partial_i T) =  0\,,\label{eq:fieldeom}\\
&   \partial_i S \partial_i T = \partial_i U\partial_i V = 0\,. \label{eq:eomxxa}\\
&    \rme^{-8b}\partial_xS\partial_x T + \rme^{2\phi}\partial_x U\partial_x V=0\,,\label{eq:eomxxb}\\
&   \rme^{-8b} (\partial_x T \partial_y S +\partial_y T\partial_x S) +  \rme^{2\phi}(\partial_x U \partial_y V + \partial_y U \partial_x V)=0\,. \label{eq:abcdcon2}
\end{align}
Note that the local Einstein constraints have the same bilinear structure as the Hanany--Witten-like consistency condition discussed in \cite{Aguilar-Gutierrez:2022kvk,Reymond:2024mwe}, although the above local, pointwise constraints on the RR fluxes are in general stronger than a relation among integrated charges. For that, let us recall how to read off the brane charges, as obtained by integrating the appropriate RR flux through the transverse cycle.  We continue to suppress the conventional numerical normalizations.  For the D($-$1) and D7 sectors our reality choice determines
\begin{equation}\label{eq:D-1D7chargeAn}
   Q_{-1} = \text{Vol}(\mathbb{T}^8)\oint \rme^{2\phi}\partial_\mu V \rmd \Sigma^\mu\, , \qquad Q_7 = \oint \rmd U\, ,
\end{equation}
and similarly for two real D3 charges, labeled by the wrapped four-torus:
\begin{equation}\label{eq:D3chargeAn}
\begin{split}
 Q_{3}^{(5678)}&=\text{Vol}(\mathbb T^4_{1234})\oint \rmd S\,,\qquad
 Q_{3}^{(1234)}=\text{Vol}(\mathbb T^4_{5678})\oint \rme^{4c-4b}\partial_\mu T\,\rmd\Sigma^\mu\,.
\end{split}
\end{equation}
In the above, $\rmd \Sigma^\mu$ is the outward-directed normal line element, while $\text{Vol}(\mathbb T^4_{1234})=\int\rmd\theta_{1234}$ denotes the coordinate volume, and similarly for the other tori. The warp factors are displayed explicitly: the factors of $a$ cancel in two dimensions, while those of $b$ and $c$ in \eqref{eq:D3chargeAn} come from the ten-dimensional Hodge star.  Thus one D3 charge is of monodromy type and the other of Gauss-law type, exactly as for the D($-$1)/D7 pair.

In special configurations in which the relevant flux components are globally aligned with a common cycle --- such as the locally codimension-one solutions of \cite{Reymond:2024mwe} --- we find:
\begin{equation}\label{constraint}
Q_{-1}Q_{7}+Q^{(1234)}_{3}Q^{(5678)}_{3}=0\,.
\end{equation}
This integrated relation is specific to the codimension-one branch. For the codimension-two branch the local equations do not by themselves fix a relation between the two charge bilinears. In the constant-dilaton sector, requiring the real part of the Killing-spinor exponent to be single-valued instead will give
\begin{equation}\label{constraint-codim2}
Q_{-1}Q_{7}-Q^{(1234)}_{3}Q^{(5678)}_{3}=0\,.
\end{equation}
The origin of this relative sign is derived in Appendix \ref{sec:charge-constraints}.


\section{New and old solutions}

We now analyse the set of equations \eqref{eq:susy1} to \eqref{eq:abcdcon2} which, given our Ansatz \eqref{eq:ansatz}, are necessary and sufficient to find a supersymmetric solution of Euclidean type IIB supergravity. We begin by checking these equations against the known brane solutions.

\subsection{Known solutions}\label{sec:known}

Our Ansatz reproduces the known single type brane solutions and bound states:

\newcommand{\knownsolrow}[3]{%
  \makebox[0.38\linewidth][l]{$\displaystyle #1$}%
  \makebox[0.23\linewidth][l]{$\displaystyle #2$}%
  \makebox[0.23\linewidth][l]{$\displaystyle #3$}%
}

\begin{enumerate}

\item D($-$1) solutions \cite{Gibbons:1995vg}:
\begin{equation}
\begin{aligned}
\knownsolrow{a=b=S=T=U=0\,,}{V=\rme^{-\phi}\,,}{\partial^2\rme^{\phi}=0\,.}
\end{aligned}
\label{eq:D-1ans}
\end{equation}

\item D7 solutions \cite{Greene:1989ya}:
\begin{equation}
\begin{aligned}
\knownsolrow{b=S=T=V=0\,,}{2a=-\phi\,,}{\partial_iU=\rme^{-\phi}\epsilon_{ij}\partial_j\phi\,,}\\[-1pt]
\knownsolrow{}{}{\partial^2\rme^{-\phi}=0\,.}
\end{aligned}
\label{eq:D7ans}
\end{equation}

\item D3 solutions \cite{Horowitz:1991cd}:
\begin{equation}
\begin{aligned}
\knownsolrow{S=U=V=0\,,}{\phi=\mathrm{ct}\,,}{4a=-4b=\log H_3\,,}\\[-1pt]
\knownsolrow{}{T=H_3^{-1}\,,}{\partial^2H_3=0\,.}
\end{aligned}
\label{eq:D3ans}
\end{equation}

\item D($-$1)-D3 bound states:
\begin{equation}
\begin{aligned}
\knownsolrow{4a=-4b=\log H_3\,,}{T=H_3^{-1}\,,}{S=U=0\,,}\\[-1pt]
\knownsolrow{V=\rme^{-\phi}\,,}{\partial^2H_3=0\,,}{\partial^2\rme^{\phi}=0\,.}
\end{aligned}
\label{eq:D1D3ans}
\end{equation}
which reduces to \eqref{eq:D3ans} when $\phi = \mathrm{ct}$ and to
\eqref{eq:D-1ans} when $H_3 = 1$. Equations
\eqref{eq:susy1}-\eqref{eq:abcdcon2} are satisfied, because they simply split
into the D($-$1) and D3 equations separately. This is the mechanism behind
the harmonic superposition rules \cite{Tseytlin:1996bh, Tseytlin:1997cs}.

\item The D3-D7 bound states similarly follow from harmonic superposition:
\begin{equation}
\begin{aligned}
\knownsolrow{4a+2\phi=-4b=\log H_3\,,}{T=H_3^{-1}\,,}{\partial_iU=\rme^{-\phi}\epsilon_{ij}\partial_j\phi\,,}\\[-1pt]
\knownsolrow{S=V=0\,,}{\partial^2H_3=0\,,}{\partial^2\rme^{-\phi}=0\,.}
\end{aligned}
\label{eq:D7D3ans}
\end{equation}
which reduces to \eqref{eq:D7ans} when $H_3=1$ and to
\eqref{eq:D3ans} when $\phi=\mathrm{ct}$.

\end{enumerate}

Finally, with all field strengths non-zero, but only allowing spatial dependence on $y$, thereby smearing the branes over $x$, leads to the D($-$1)-D3-D7 solution
of \cite{Reymond:2024mwe}. Its explicit form will follow directly from the general local solutions below.

\subsection{General local solutions}\label{sec:local-solutions}
The holomorphic functions in \eqref{eq:susy1} already contain most of the local information.  Their imaginary parts are harmonic functions, which we denote by
\begin{equation}\label{eq:H12}
    H_1=V-\rme^{-\phi}=\Im\tau_1\,,\qquad
    H_2=T-\rme^{4b}=\Im\tau_2\,.
\end{equation}
These are the harmonic conjugates of $U=\Re \tau_1$ and $S=\Re \tau_2$, respectively.  On patches where $\rmd H_1$ and $\rmd H_2$ do not vanish,\footnote{If one of the $H_i$ is constant and cannot serve as a local coordinate, then we find that either $\rme^\phi$ or $\rme^{-4b}$ is harmonic, and we are back with brane solutions already discussed in section \ref{sec:known}.} the orthogonality relations \eqref{eq:eomxxa} further imply that $\phi$ depends only on $H_1$ and $b$ only on $H_2$. Their dependence is fixed by the flux equations \eqref{eq:fieldeom}:
\begin{equation}\label{eq:riccati}
    \frac{\rmd \rme^{\phi}}{\rmd H_1}=\rme^{2\phi}+c_1\,,\qquad
    \frac{\rmd \rme^{-4b}}{\rmd H_2}=\rme^{-8b}+c_2\,,
\end{equation}
where $c_1$ and $c_2$ are real constants. For positive $c_i$, the solutions to these Riccati equations are trigonometric, while for negative $c_i$ they are hyperbolic. For example, the first equation in \eqref{eq:riccati} gives
\begin{equation}
\rme^\phi=
\begin{cases}
\sqrt{c_1}\,\tan\!\left[\sqrt{c_1}(H_1-h_1)\right], & c_1>0\,,\\[2mm]
\sqrt{-c_1}\,\tanh\!\left[\sqrt{-c_1}(h_1-H_1)\right], & c_1<0\,,\\[1mm]
\sqrt{-c_1}\,\coth\!\left[\sqrt{-c_1}(h_1-H_1)\right], & c_1<0\,,\\[2mm]
\dfrac{1}{h_1-H_1}, & c_1=0\,.
\end{cases}
\label{eq:phi-solutions}
\end{equation}
Here $h_1$ is an integration constant. For $c_1<0$ there is also the constant fixed-point solution $\rme^\phi=\sqrt{-c_1}$ which separates the two hyperbolic branches. The branches above are clearly related by analytic continuation. The given expressions are understood on connected intervals where $\rme^\phi>0$. The $b$ sector is identical after replacing $(\phi,H_1,c_1)$ by $(-4b,H_2,c_2)$.  

The two sectors are tied together by the two remaining Einstein constraints \eqref{eq:eomxxb} and \eqref{eq:abcdcon2}, which using \eqref{eq:riccati} can be written very compactly as the single complex equation
\begin{equation}\label{eq:local-constraint}
    c_1(\partial_z H_1)^2+c_2(\partial_z H_2)^2=0\,,\qquad
    \partial_z=\tfrac12(\partial_x-\rmi\partial_y)\,.
\end{equation}

If $c_1$ and $c_2$ have the same sign, $H_1$ is proportional to the harmonic conjugate of $H_2$. This is a genuinely codimension-two solution. We discuss examples in section \ref{sec:cylindrical-limits}.  

When $c_1$ and $c_2$ instead have opposite signs, $H_1$ and $H_2$ are proportional and the solution reduces to codimension one, at least locally. 

If either $c_1$ or $c_2$ vanishes, the dilaton solution \eqref{eq:phi-solutions} or its torus-size analog says $\rme^{-\phi}$ or $\rme^{4b}$ is harmonic. This brings us back to one of the already known solutions of section \ref{sec:known}.

Finally, the transverse warp factor $a$ is fixed by the remaining BPS equation \eqref{eq:gravi2b}. Written in terms of $H_{1,2}$, it determines
\begin{equation}\label{eq:warpfactor-quadrature}
    \partial_i(4a)=-2\rme^\phi\partial_i H_1-4\partial_i b-2\rme^{-4b}\partial_i H_2+8\epsilon_{ij}\partial_j\Im f\,.
\end{equation}
For $\Im f=0$ this can be immediately integrated, as we do explicitly for the solutions below.

Equations \eqref{eq:riccati}, \eqref{eq:local-constraint} and \eqref{eq:warpfactor-quadrature}, together with the holomorphic data $\tau_{1,2}$ and the harmonic choice of $\Im f$, therefore provide a local classification of the supersymmetric solutions within our Ansatz. Apart from the degenerate limits that reproduce the known branes, the sign of $c_1c_2$ separates the genuinely codimension-two branch (conjugate harmonics) from the locally codimension-one branch (proportional harmonics).

\paragraph{Locally codimension-one solution.}
In the case where $c_1$ and $c_2$ in \eqref{eq:riccati} have opposite signs, $H_1$ and $H_2$ are locally proportional to each other. One may call the common harmonic coordinate $y$ and its conjugate $x$.  Writing then
\begin{equation}
    U=\alpha x\,,\qquad S=\gamma x\,,\qquad
    \rme^{2\phi}\partial_yV=\beta\,,\qquad
    \rme^{-8b}\partial_yT=\delta\,,
\end{equation}
with constant $\alpha,\beta,\gamma,\delta$, the constraint \eqref{eq:local-constraint} has become
\begin{equation}\label{eq:con}
    \alpha\beta+\gamma\delta=0\,.
\end{equation}
The complete solution, given the Ansatz \eqref{eq:ansatz}, can be summarized as
\begin{equation}\label{eq:codim1-full}
\begin{aligned}
\rme^{-\phi}&=-\sqrt{\frac{\alpha}{\beta}}\,
\tanh\!\left[\sqrt{\alpha\beta}\,(y-y_0)\right],
\qquad V=\alpha y+\rme^{-\phi},\\[1mm]
\rme^{4b}&=-\sqrt{\frac{\gamma}{\delta}}\,
\tanh\!\left[\sqrt{\gamma\delta}\,(y-y_1)\right],
\qquad T=\gamma y+\rme^{4b},\\[1mm]
a&=a_0+\frac12\log\left|\sinh\!\left(\sqrt{\alpha\beta}(y-y_0)\right)\right|
+\frac14\log\left|\sinh\!\left(2\sqrt{\gamma\delta}(y-y_1)\right)\right|.
\end{aligned}
\end{equation}
Here $c=-b$, and additive constants in $U,V,S,T$ have been suppressed.  

Since $\alpha\beta+\gamma\delta=0$, one of the hyperbolic functions is analytically continued to a trigonometric branch.  
This is precisely the partially smeared D($-$1)/D7 solution of \cite{Reymond:2024mwe}, now obtained directly as the codimension-one branch of the two-dimensional equations.

Equivalently, for a radially symmetric solution, replace $y = -\log r$ and $x = \theta$. Hence, the same solution can also be interpreted as a codimension-two solution on a punctured plane. The charges \eqref{eq:D-1D7chargeAn} and \eqref{eq:D3chargeAn}, evaluated by integrating around $r=0$, are then given by
\begin{equation}\begin{split}
    Q_7 =2\pi \alpha \, &, \quad Q_{-1} =-2\pi  \beta {\rm Vol}(\mathbb T^8)\,,\\
    Q_{3}^{(1234)}=-2\pi\delta\,{\rm Vol}(\mathbb T^4_{5678})\,&,\quad Q_{3}^{(5678)}=2\pi\gamma \,{\rm Vol}(\mathbb T^4_{1234})\,.\end{split}
\end{equation}
Consequently, the constraint \eqref{eq:con} immediately implies
\begin{equation}\label{eq:Qrelation}
    Q_{-1}Q_7+ Q_{3}^{(1234)}Q_{3}^{(5678)}=0\,.
\end{equation}

\subsection{Meromorphic solutions and near-horizon limits}\label{sec:cylindrical-limits}
Consider now genuinely codimension-two solutions. As argued below \eqref{eq:local-constraint}, this requires $H_1$ to be proportional to the harmonic conjugate of $H_2$. Equivalently, $\tau_1$ must be proportional to $\rmi\tau_2$, up to a constant which we can set to zero in what follows.

\subsubsection{Constant dilaton and torus size}\label{sec:constant-scalars}
The simplest solutions of this kind have constant $\phi=\phi_0$ and $b=b_0$.  Then \eqref{eq:local-constraint} implies
\begin{equation}\label{eq:constant-relation}
  \rme^{\phi_0}\tau_1 = \sigma\, \rmi\rme^{-4b_0}\tau_2\,,\qquad \sigma=\pm1\,.
\end{equation}
 Integrating \eqref{eq:warpfactor-quadrature}, with the spinor phase $\Im(f)$ set to zero,  determines the scale factor of the transverse metric. In terms of
\begin{equation}\label{eq:R-def}
    R(z)=\exp\!\left(\frac{\sigma+\rmi}{2}\,\rme^{\phi_0}\tau_1(z)\right),\qquad |R(z)|^2 = \exp\!\left(-\rme^{\phi_0}H_1 - \rme^{-4b_0}H_2\right)\,,
\end{equation}
the final solution takes the particularly simple form
\begin{align}\label{eq:constant-solution}
    \rmd s_E^2={}&|R(z)|^2\rmd z\rmd\bar z
    +\rme^{2b_0}\sum_{i=1}^{4}\rmd\theta_i^2
    +\rme^{-2b_0}\sum_{i=5}^{8}\rmd\theta_i^2\,,\qquad \epsilon_z = \rme^{-\phi_0/8}|R(z)|^{-1/2}\,\epsilon_0\,.\\
    F_1={}&\rmd\tau_1\,,\qquad
    \mathcal F=-\sigma\rmi\,\rme^{\phi_0+4b_0}
    \rmd\theta_{1234}\wedge\rmd\tau_1\,,\qquad
    F_5=(1-\rmi\star)\mathcal F\,.
\end{align}
Thus any holomorphic $\tau_1$ determines a locally flat transverse metric through the holomorphic function $R(z)$, whose zeros, poles and monodromies contain the global information. 


\paragraph{A linear example}
A useful regular example is $\tau_1=qz$ with $q\in\mathbb R_0$.
Then we have 
\begin{equation}
    R(z)=\exp\!\left(\frac{\sigma+\rmi}{2}\,\rme^{\phi_0}qz\right),
    \qquad F_1=q\,\rmd z\,,
\end{equation}
or, in terms of the original variables
\begin{equation}
    \rmd U=q\,\rmd x\,,\qquad
    \rmd V=q\,\rmd y,\qquad \rmd S=\sigma q\rme^{\phi_0+4b_0}\rmd y\,,
    \qquad \rmd T=-\sigma q \rme^{\phi_0+4b_0}\rmd x\,.
\end{equation}
On the unquotiented $z$-plane there is no non-trivial one-cycle, and all four charges vanish. However, the metric is invariant under shifts $\Delta z\, \propto \; 1+\rmi \sigma$. The simplest example is such that $R$ itself does not change:
\begin{equation}
    \Delta z=\frac{2\pi\rme^{-\phi_0}}{q}(1+\rmi\sigma)\,,
\end{equation}
we can identify $z\sim z +\Delta z$, leading to a non-trivial cycle and charges \eqref{eq:D-1D7chargeAn} and \eqref{eq:D3chargeAn}:
\begin{equation}\begin{split}\label{eq:ex1charges}
    Q_7    =2\pi\rme^{-\phi_0}\,&,   \quad     Q_{-1} =-2\pi\rme^{\phi_0}{\rm Vol}(\mathbb T^8)\,,\\
    Q_{3}^{(5678)} =2\pi\rme^{4b_0}{\rm Vol}(\mathbb T^4_{1234})\,&,\quad Q_{3}^{(1234)} =-2\pi\rme^{-4b_0}{\rm Vol}(\mathbb T^4_{5678})\,.
    \end{split}
\end{equation}
Note that the $q$-dependence has dropped out. Moreover, the charges satisfy 
\begin{equation}
Q_{-1}Q_7 = Q_{3}^{(1234)} Q_{3}^{(5678)}\,.
\end{equation}
Charge quantization is compatible with this relation. Restoring the conventional RR normalizations, the elementary charge units obey $q_{-1}q_7=q_3^2=(2\pi)^8\alpha'^4$. Hence, writing $Q_p=q_pN_p$, the bilinear constraint reduces simply to
\begin{equation}
    N_{-1}N_7=N_3^{(1234)}N_3^{(5678)}\,,
\end{equation}
with no residual factors of $2\pi$ or $\alpha'$ obstructing integrality.
Note the relative minus sign as compared to the codimension-one solution  \eqref{eq:Qrelation}. Both are of course compatible with the local equations \eqref{eq:eomxxb} and \eqref{eq:abcdcon2}, but they arise differently: \eqref{eq:Qrelation} follows from the local constraint \eqref{eq:con} alone, while here the local equations leave the two bilinears unrelated and the relative sign is fixed by global supersymmetry. We give the derivation in Appendix \ref{sec:charge-constraints}.

Generalizations of the linear example to $\tau \sim z^n$, or even generic polynomials without linear term, lead to local solutions but without a natural way to obtain brane charges. More physical examples can be found using logarithms, which we discuss next.

\paragraph{Logarithmic example} Another solution arises when considering a simple pole in $R(z)$:
\begin{equation}\label{eq:simple-pole}
    R(z)=\frac{L}{z-z_0}\,,\qquad
    \tau_1=\rme^{-\phi_0}(\sigma-\rmi)\log\!\frac{L}{z-z_0}\,.
\end{equation}
Writing $z-z_0=r\rme^{\rmi\theta}$ and $\rho=\log r$ then gives
\begin{equation}\label{eq:cylinder}
    \rmd s_{E,2}^2=L^2\left(\frac{\rmd r^2}{r^2}+\rmd\theta^2\right)
    =L^2(\rmd\rho^2+\rmd\theta^2)\,,\qquad \theta\sim\theta+2\pi\,.
\end{equation}
This exact cylindrical throat seems a promising candidate for the $\mathbb R\times S^1$ part of the D($-$1)/D7 near-horizon geometry. The fluxes are constant in $(\rho,\theta)$ coordinates, and the charges are the same as \eqref{eq:ex1charges}, up to an overall minus sign. However, the full solution differs from the previous example. They are instead related via \eqref{eq:harm}.

\paragraph{Several logarithmic contributions}

By adding a constant to the above $R(z)$, we get
\begin{equation}\label{eq:R-completion}
    R(z)=\frac{z-z_1}{z-z_0}\,.
\end{equation}
This keeps the cylindrical throat near $z_0$, while the large-$|z|$ transverse metric asymptotes to the Euclidean plane. The zero at $z=z_1$ is a compensating excess-angle defect. Its microscopic source is part of the global completion problem that we will not address here.\footnote{It is useful to compactify the transverse space, so that $\omega=R(z)\,\rmd z$ can be regarded as a meromorphic one-form on $S^2$. For the simple-pole solution, $\omega=L\,\rmd z/(z-z_0)$ has simple poles at $z=z_0$ and $z=\infty$, giving total divisor degree $-2$, as required for any meromorphic one-form on $S^2$. The physical transverse space is obtained by removing these two points and is therefore $S^2\setminus\{z_0,\infty\}\simeq\mathbb C^\ast\simeq\mathbb R\times S^1$, with Euler characteristic $\chi=0$. For the solution \eqref{eq:R-completion}, $\omega$ instead has a simple zero at $z=z_1$, a simple pole at $z=z_0$, and a double pole at infinity. Its divisor again has degree $1-1-2=-2$. Yet, the geometry of the two ends is different: $z_0$ is cylindrical while $z=\infty$ is asymptotically planar, and the zero at $z=z_1$ gives a $4\pi$ cone point. 
}

More generally, logarithmic contributions to $\tau_1$ can be superposed. For example, let
\begin{equation}
R(z)=\prod_{j=1}^{N} \left(z-z_j\right)^{m_j}\,,
\qquad \tau_1=\rme^{-\phi_0}(\sigma-\rmi)\log R(z)\,,
\end{equation}
where the $z_j$ are distinct, and $m_j\in\mathbb{Z}_0$. Away from the zeros and poles, these data determine a solution through \eqref{eq:constant-solution}, with
\begin{equation}
    F_1=\rme^{-\phi_0}(\sigma-\rmi)
\sum_{j=1}^{N}m_j\frac{\rmd z}{z-z_j}\,.
\end{equation}
Moreover, for a positively oriented loop $\gamma_j$ around $z_j$,
\begin{equation}
    \oint_{\gamma_j}F_1 =2\pi\rmi\,\rme^{-\phi_0}(\sigma-\rmi)m_j\,,
\end{equation}
with the corresponding five-form periods fixed by \eqref{eq:constant-solution}.

Near $z_j$, writing $R(z)\sim A_j(z-z_j)^{m_j}$ and $z-z_j=r\rme^{\rmi\theta}$, the transverse metric behaves as
\begin{equation}
    \rmd s_{E,2}^2\sim |A_j|^2r^{2m_j} \left(\rmd r^2+r^2\rmd\theta^2\right)\,,
\qquad \theta\sim\theta+2\pi\,.
\end{equation}
The periodicity of $\theta$ is enforced by having a consistent spin structure.
A simple pole, $m_j=-1$, therefore gives an asymptotically cylindrical end, with circumference $2\pi|A_j|$. A pole of order $p\geq2$, namely $m_j=-p$, gives a cone of opening angle $2\pi(p-1)$ at infinite distance. A zero of order $m_j\geq1$ instead gives a conical singularity of opening angle $2\pi(m_j+1)$ at finite distance.

\subsubsection{More general solutions}\label{sec:more-general-solutions}
The local BPS equations admit more general solutions than the constant-phase, constant-scalar configurations discussed above. We explain in Appendix~\ref{sec:harmonic-phase} how to include a non-zero harmonic Killing-spinor phase $\Im f$. In Appendix~\ref{sec:general-solutions-appendix}, we show how to let the dilaton or relative torus size run according to \eqref{eq:riccati}. 

Although these configurations solve the local supersymmetry equations, their global
interpretation is less satisfactory: angular periodicity of the metric can be restored only at the price of a non-trivial angular monodromy of the Killing spinor, which generically also changes its norm. Hence the Killing spinor is not globally single-valued. We therefore regard these configurations as useful local extensions of the solution space, rather than additional global D($-1$)/D7 backgrounds.

\section{Conclusions}
We analyzed supersymmetric type IIB configurations carrying D($-$1), D7 and two complementary Euclidean D3 fluxes with full dependence on the two transverse directions.  Holomorphicity organizes the local solution in terms of two real harmonic functions $H_1$ and $H_2$, which determine the dilaton $\phi$ and relative torus size $b$ through Riccati equations \eqref{eq:riccati}. 

The $\phi$ and $b$ sectors are tied together through the local Einstein constraints \eqref{eq:local-constraint}. This gives a local classification within our Ansatz. In particular, the relative sign of the two integration constants in \eqref{eq:riccati} appears in \eqref{eq:local-constraint} and determines whether the solution is genuinely codimension two (the harmonics are conjugate) or reduces locally to the smeared solution of~\cite{Reymond:2024mwe} (the harmonics are proportional). In section \ref{sec:constant-scalars}, we discussed representatives of the former case, in particular those most relevant for the D($-$1)/D7 interpretation, while the solution for the latter case was given explicitly in \eqref{eq:codim1-full}.

The local Einstein constraints tie the four flux sectors in a bilinear way, but their implication for integrated charges is branch-dependent. For the proportional, locally codimension-one branch they imply
\begin{equation}
    Q_{-1}Q_{7}+ Q^{(1234)}_{3}Q^{(5678)}_{3}=0\,,
\end{equation}
whereas in the conjugate branch, if the dilaton and torus size are constant, the local equations leave the two bilinears unrelated and global supersymmetry instead imposes
\begin{equation}
    Q_{-1}Q_{7}- Q^{(1234)}_{3}Q^{(5678)}_{3}=0\,.
\end{equation}
The distinction is derived in Appendix \ref{sec:charge-constraints}.

The most interesting codimension-two sector has constant dilaton and constant torus size.  Its transverse metric is encoded by the single holomorphic function $R(z)$ in \eqref{eq:R-def}.  A simple pole gives an exact $\mathbb R\times S^1$ cylinder, while \eqref{eq:R-completion} provides an asymptotically flat completion with a controlled near-core limit. More general local solutions, including harmonic dressings with non-zero $\Im f$ and running-modulus deformations, are collected in Appendix \ref{sec:general-solutions-appendix}; their natural angular identification leads to a non-trivial Killing-spinor monodromy (including a monodromy of its norm).

The simple-pole throat solution is a direct candidate for the formal $p=-1$ entry in the following family of $\AdS$ solutions:
\begin{equation}
    \mathrm{AdS}_{p+2}\times S^{p+2}\times\mathbb T^{6-2p}\,.
\end{equation}
The $p=-1$ entry of course degenerates, so that ``$\mathrm{AdS}_1$'' is simply the $\mathbb R$ factor appearing in our cylinder solution, rather than a space of non-zero intrinsic curvature.

It is natural to ask whether this background admits the proposed ``${\rm CFT}_0$'' interpretation in terms of the D($-$1)/D7 matrix model of \cite{Billo:2021xzh}, possibly as a fixed point of a Br\'ezin--Zinn-Justin matrix RG flow \cite{Brezin_ZJ_1992,Zinn-Justin:2014wva, Guan:2026sjw}.  The simultaneous electric and magnetic RR data may also point to an ensemble interpretation in which charge sectors are summed over \cite{Aharony:2024fid,Gautason:2025plx}.  


\subsection*{Acknowledgments}
We are grateful to Lander Umans for initial work on this topic \cite{lander}. We benefited from discussions with Iosif Bena, Dimitrios Toulikas and Laurent Gallot and from earlier work with Sergio Aguilar, Mario Trigiante and Sebastien Reymond. 
The work of GC and TVR is supported by the KU Leuven C1 grant ZKE7799C16/25/010. KP was supported by the STFC consolidated grant ST/T000694/1 and Simons Investigator award \#620869.

\appendix

\section{Equations of motion}\label{sec:EOM}
The equations of motion of type IIB string theory are well-known and appear in many places, see for instance \cite{Becker:2006dvp}. We start from the following Einstein-frame Lagrangian:
\begin{equation}
   \mathcal{L} = \sqrt{g}(R- \frac12 (\nabla \phi)^2 -\frac12\rme^{2\phi}F^2_1- \frac{1}{4\cdot 5!}F^2_5)\,.
\end{equation}
Importantly, self-duality of $F_5$ must be imposed separately. In the above $F^2_p = F_{\mu_1 \dots \mu_p}F^{\mu_1 \dots \mu_p}$, without complex conjugation.
The $\phi$ and $C_0$ equations of motion then each contain a real and imaginary part, leading to four equations, which in our Ansatz \eqref{eq:ansatz} become
\begin{align}
&  \partial^2\phi + \rme^{2\phi}((\partial V)^2-(\partial U)^2) + 4\partial \phi \partial(b+c)  =0 \,, \label{eq:dil1} \\
&    \partial U \partial V = 0\,. \label{eq:dil2}\\
  & \partial(\rme^{4(b+c) + 2\phi}\partial U)=\partial(\rme^{4(b+c) + 2\phi}\partial V)=0\,. \label{eq:F1eom}
\end{align}
Here and in what follows, $\partial^2$ simply means $\partial^2_x + \partial^2_y$, in order to reduce clutter. Similarly, $\partial U \partial V$ is shorthand for $\partial_x U \partial_x V + \partial_y U \partial_y V$, without additional metric factors.

For the 5-form we need to impose the Bianchi identity --- the equation of motion then follows from self-duality. Our Ansatz implies $\rmd F_5 = \rmd (1-\rmi \star)\cal{F}= -\rmi \,\rmd \star \cal{F}$. The resulting equations are
\begin{equation}\label{eq:F5bianchi}
    \partial\!\left(\rme^{4c-4b}\partial S\right)=0\,,\qquad
    \partial\!\left(\rme^{4c-4b}\partial T\right)=0\,.
\end{equation}

Finally, the Einstein equations 
\begin{equation}
    R_{\mu\nu} = \frac12 \partial_\mu \phi\partial_\nu \phi + \frac12\rme^{2\phi} F_{1\mu}F_{1\nu}+\frac{1}{4\cdot 4!}F_{5\mu ijkl}F_{5\nu}^{ijkl}\,,
\end{equation}
split into ten separate equations, five real and five imaginary. The $\theta\theta$-components imply
\begin{align}
& R_{\theta_5\theta_5}:\quad   (\partial S)^2-(\partial T)^2 - 4\rme^{8b}\partial^2c -16\rme^{8b} \partial c\partial(b+c)= \partial S \partial T=0\,,\label{eq:reom1}\\
& R_{\theta_1\theta_1}:\quad (\partial T)^2-(\partial S)^2 - 4\rme^{8b}\partial^2b -16\rme^{8b} \partial b\partial(b+c)= \partial S \partial T=0\,.
\end{align}
Combining these two sets of equations together, we find the condition
\begin{equation}\label{eq:Reom4}
    \partial^2\rme^{4(b+c)}=0\,.
\end{equation}
Next, the $xx$-component ($yy$-components are found by $x\leftrightarrow y$) yields
\begin{equation}
\begin{split}
    &-2(\partial_x \phi)^2+2\rme^{2\phi}((\partial_x V)^2-(\partial_x U)^2)+\rme^{-8b}((\partial_x T)^2-(\partial_x S)^2+(\partial_y S)^2-(\partial_y T)^2)\\
    &+16\partial_x a \partial_x(b+c)-16\partial_y a \partial_y(b+c)-4\partial^2 a-16\partial^2_x(b+c)-16((\partial_x b)^2+(\partial_xc)^2)=0\, ,\label{eq:Reomxx}
    \end{split}
\end{equation}
as well as 
\begin{equation}
   \partial_x S \partial_x T - \partial_y S \partial_y T + 2\rme^{8b + 2\phi}\partial_x U \partial_x V=0 \,.\label{eq:Ieomxx}
\end{equation}
Finally, the off-diagonal $xy$-component tells us that
\begin{equation}\label{eq:reom7}\begin{split}
   &-\partial_x\phi\partial_y\phi + \rme^{2\phi}(\partial_xV\partial_y V-\partial_xU\partial_y U ) + \rme^{-8b}(\partial_xT\partial_y T-\partial_xS\partial_y S)\\&+8\partial_x a \partial_y(b+c) + 8 \partial_y a \partial_x(b+c)-8 \partial_{xy}(b+c) -8 \partial_x c \partial_y c - 8 \partial_x b \partial_y b=0\,,
   \end{split}
\end{equation}
and
\begin{equation}\label{eq:reom8}
    \partial_x T \partial_y S +\partial_y T\partial_x S +  \rme^{8b + 2\phi}(\partial_x U \partial_y V + \partial_y U \partial_x V)=0\,.
\end{equation}

\section{Which equations of motion are implied?}\label{sec:implied}
We now ask which of the equations of motion in Appendix \ref{sec:EOM} are implied by the BPS condition of section \ref{sec:susy}. The result of the analysis below is summarized in section \ref{sec:bps2}.

The gravitino variation \eqref{eq:gravi3} imposes $c=-b+\mathrm{ct}.$, which simplifies many equations. Since the constant drops out of equations, we can set it to zero in practice. This is equivalent to absorbing it in the circle coordinates $\theta$. 

Then, using $c=-b$ together with the dilatino variation \eqref{eq:dilatino} (which implies in particular $\partial^2(V-\rme^{-\phi})=0$), we see that the real part \eqref{eq:dil1} of the dilaton equation of motion is already implied by the second $F_1$ equation in \eqref{eq:F1eom}, namely
\begin{equation}\label{eq:phiV}
     \partial(\rme^{2\phi}\partial V) =0\,.
\end{equation}
This equation by itself does not follow from the BPS equations.

On the other hand, the first $F_1$ equation of motion in \eqref{eq:F1eom} becomes $\partial \phi \partial U = 0$, which is implied by the dilaton equation of motion \eqref{eq:dil2}:
\begin{equation}\label{eq:UV}
    \partial U \partial V =0\,,
\end{equation}
which must be also imposed separately. The equivalent relation for $S, T$ follows from the $\theta\theta$-component \eqref{eq:reom1} of the Einstein equation:
\begin{equation}\label{eq:ST}
\partial S \partial T = 0\,,
\end{equation}
which does not follow from the BPS conditions either.


Next, after applying $c=-b$, the Bianchi equations \eqref{eq:F5bianchi} reduce to $\partial(\rme^{-8b}\partial S)=0$ and $\partial(\rme^{-8b}\partial T)=0$. The first one already follows from the gravitino variation \eqref{eq:gravi4} together with \eqref{eq:ST}. This leaves only 
 \begin{equation}\label{eq:bT}
     \partial(\rme^{-8b}\partial T)=0
 \end{equation}
 as an independent condition, mirroring \eqref{eq:phiV}.
 
Now we look at the remaining Einstein equations, using $c=-b$ as well as 
the BPS holomorphicity conditions \eqref{eq:hol1} and \eqref{eq:hol2}. It turns out that the remaining real $\theta\theta$-components are already implied by the $F_5$ Bianchi \eqref{eq:F5bianchi}. The imaginary part \eqref{eq:Ieomxx} of the $xx$-equation becomes
\begin{equation}\label{eq:abcdcon}
   \partial_x S \partial_x T  + \rme^{8b + 2\phi}\partial_x U \partial_xV =0 \,.
\end{equation}
and similarly for the $yy$-component, which then already follows from \eqref{eq:UV} and \eqref{eq:ST} together with the above $xx$-component. This now implies the real part \eqref{eq:reom7} of the $xy$-component, while its imaginary part \eqref{eq:reom8} 
\begin{equation}\label{eq:abcdconI}
    \partial_x T \partial_y S +\partial_y T\partial_x S +  \rme^{8b + 2\phi}(\partial_x U \partial_y V + \partial_y U \partial_x V)=0\,,
\end{equation}
provides an additional constraint. Finally, the sum of the real parts of $xx$- and $yy$-components follows from taking the divergence of \eqref{eq:gravi2}, while their difference is implied by \eqref{eq:abcdconI}.

As a final comment, note that the gravitino equation \eqref{eq:gravi2b} does impose an integrability condition, which we have not quite taken into account yet: 
\begin{equation}\label{eq:int}
    \rme^\phi (\partial_x \phi \partial_y V -\partial_y \phi \partial_x V)  - 4\rme^{-4b}(\partial_x b \partial_y T - \partial_y b \partial_x T)+ 4 \partial^2 \Im(f)=0\,.
\end{equation}
Here $f$ determines the spatial dependence of the spinor through $\epsilon = \rme^{f}\epsilon_0$. Together with the holomorphicity conditions \eqref{eq:hol1} and \eqref{eq:hol2}, this can be rewritten as
\begin{equation}\label{eq:int2}
 \rme^{-8b}\partial S \partial T   + \rme^{2\phi}\partial U \partial V = 4\partial^2 \Im(f)\,.
\end{equation}
Moreover, \eqref{eq:gravi1b} implies that $\Im(f)$ must be harmonic. This means that when the BPS equations hold, it suffices to impose either \eqref{eq:UV} or \eqref{eq:ST}, since the other then follows from the above integrability condition.

\section{Charge constraints}\label{sec:charge-constraints}
Here we derive the integrated charge relations on the two different branches of the local solution space. From \eqref{eq:H12} and the Riccati equations \eqref{eq:riccati} one finds
\begin{equation}\label{eq:appendix-charge-differentials}
    \rme^{2\phi}\rmd V=-c_1\rmd H_1\,,\qquad
    \rme^{-8b}\rmd T=-c_2\rmd H_2\,.
\end{equation}
Holomorphicity of $\tau_1=U+\rmi H_1$ and $\tau_2=S+\rmi H_2$ gives, with our orientation conventions,
\begin{equation}
    \star\rmd H_1=-\rmd U\,,\qquad
    \star\rmd H_2=-\rmd S\,.
\end{equation}
Then, using the charge definitions \eqref{eq:D-1D7chargeAn} and \eqref{eq:D3chargeAn}, for a common closed transverse cycle~$\gamma$,
\begin{equation}\label{eq:appendix-charge-products}
\begin{split}
    Q_{-1}Q_7
    &=c_1\,\mathrm{Vol}(\mathbb T^8)
      \left(\oint_\gamma \rmd U\right)^2\,,\\
    Q_{3}^{(1234)}Q_{3}^{(5678)}
    &=c_2\,\mathrm{Vol}(\mathbb T^8)
      \left(\oint_\gamma \rmd S\right)^2\,.
\end{split}
\end{equation}

On the proportional branch we may write $\tau_1=k\tau_2+\mathrm{ct.}$ with real $k$. The local constraint \eqref{eq:local-constraint} gives $k^2c_1=-c_2$, while $\rmd U=k\rmd S$. Then \eqref{eq:appendix-charge-products} immediately yields
\begin{equation}
    Q_{-1}Q_7+Q_{3}^{(1234)}Q_{3}^{(5678)}=0\,.
\end{equation}
This is the relation obeyed by the locally codimension-one solutions.

On the conjugate branch we can instead write $\tau_1=\rmi k\tau_2+\mathrm{ct.}$, again with real $k$. In this case, \eqref{eq:local-constraint} gives $k^2c_1=c_2$. Since
\begin{equation}
    \rmd U=-k\rmd H_2\,,\qquad \rmd H_1=k\rmd S\,,
\end{equation}
we obtain
\begin{equation}\label{eq:appendix-conjugate-bilinears}
\begin{split}
    Q_{-1}Q_7
    &=c_2\,\mathrm{Vol}(\mathbb T^8)
      \left(\oint_\gamma \rmd H_2\right)^2\,,\\
    Q_{3}^{(1234)}Q_{3}^{(5678)}
    &=c_2\,\mathrm{Vol}(\mathbb T^8)
      \left(\oint_\gamma \rmd S\right)^2\,.
\end{split}
\end{equation}
Thus the local equations alone impose neither the sum nor the difference of the two charge bilinears to vanish.

The additional condition relevant for a global supersymmetric solution follows directly from the gravitino equations. Eliminating the transverse conformal factor and the Killing-spinor phase between \eqref{eq:gravi1b} and \eqref{eq:gravi2b} gives
\begin{equation}\label{eq:appendix-ref-f}
    \rmd\Re(f)=\frac18\rmd\phi-\frac12\rmd b
    +\frac14\left(\rme^\phi\rmd V+\rme^{-4b}\rmd T\right)\,.
\end{equation}
For a globally defined Killing-spinor norm, $\Re(f)$ must have no monodromy. If the scalar fields are single-valued around $\gamma$, integration of \eqref{eq:appendix-ref-f} therefore requires
\begin{equation}\label{eq:appendix-global-susy}
    \oint_\gamma\left(\rme^\phi\rmd V+\rme^{-4b}\rmd T\right)=0\,.
\end{equation}
For the constant-scalar conjugate branch, $c_1=-\rme^{2\phi_0}$ and $c_2=-\rme^{-8b_0}$, while \eqref{eq:constant-relation} says
\begin{equation}
    k=\sigma\rme^{-\phi_0-4b_0}\,.
\end{equation}
Using \eqref{eq:appendix-charge-differentials}, condition \eqref{eq:appendix-global-susy} becomes
\begin{equation}
    \rme^{\phi_0}\oint_\gamma\rmd H_1
    +\rme^{-4b_0}\oint_\gamma\rmd H_2=0
    \quad\Longrightarrow\quad
    \oint_\gamma\rmd H_2=-\sigma\oint_\gamma\rmd S\,.
\end{equation}
Substituting this into \eqref{eq:appendix-conjugate-bilinears} gives
\begin{equation}
    Q_{-1}Q_7-Q_{3}^{(1234)}Q_{3}^{(5678)}=0\,.
\end{equation}
Hence the relative sign between the two integrated charge constraints is not a convention: the plus sign follows locally on the proportional branch, whereas on the constant-scalar conjugate branch the minus sign follows only after imposing the global no-monodromy condition on the Killing-spinor norm.

\section{Solutions with non-zero Killing-spinor phase}\label{sec:harmonic-phase}
So far, we considered solutions whose Killing spinor has a constant phase: $\epsilon = \rme^f \epsilon_0$ with vanishing $\Im f$. This was particularly relevant when integrating the gravitino variation \eqref{eq:gravi2b} to find the conformal factor $a$, leading to the locally flat solutions \eqref{eq:constant-solution}. For these solutions, we can trivialize the metric by a holomorphic change of transverse coordinate $z=F(z')$.  Writing $F'(z') = |F'(z')| e^{i\alpha(z')}$ makes clear that the holomorphic coordinate transformation induces a rescaling of the warp factor and a local rotation of the
orthonormal frame by the angle $\alpha$. Since spinors transform as
$\epsilon\to e^{\frac{\alpha}{2}\Gamma_{xy}}\epsilon$, using $\Gamma_{xy}\epsilon=i\epsilon$, we find:
\begin{align}
&    a(z,\bar z) \ \to\  a(F,\bar F)+\log|F'(z')|\,, \label{eq:warp-coordinate-shift}\\
&   f\ \to\ \ f+\frac{\rmi}{2}\arg F'\,. \label{eq:spinor-coordinate-shift}
\end{align}
In particular, the metric \eqref{eq:constant-solution} becomes trivial when choosing $F' = R^{-1}$.

More generally, the equations are invariant under shifting $a$ by a harmonic function $h$ and simultaneously shifting $\Im f$ by its harmonic conjugate $\tilde{h}$:
\begin{equation}\label{eq:harm}
    a\ \to\  a+h\,,\quad f \ \to\  f + \frac{\rmi}{2}\,\widetilde h\,,
\end{equation}
where
\begin{equation}
    \partial^2h=0\,, \quad \partial_i h=\epsilon_{ij}\partial_j\widetilde h\,.
\end{equation}
Such a ``harmonic dressing'' is not generically generated by a global coordinate transformation (unlike what we did in \eqref{eq:warp-coordinate-shift}, \eqref{eq:spinor-coordinate-shift}) . It changes the Killing-spinor phase but leaves its real part unchanged.

Thus we have a family of BPS harmonic dressings with identical $\phi,b,U,V,S,T$. In the constant dilaton and torus examples considered up to now, we chose the representative $\Im f=0$. More generally, a particular representative can be selected by imposing for instance an angular isometry and finite asymptotic circle size. Below, we give an example of this.  

\paragraph{From the linear example to the cylinder}
As a concrete example, we can see how the linear and logarithmic examples of section \ref{sec:constant-scalars} are related via \eqref{eq:harm}.

Let us start with the linear example $\tau_1=qz$. Introduce coordinates $t=u+\rmi v$ and a positive constant $\ell$ by 
\begin{equation}
    t=\frac{\sigma+\rmi}{2}\rme^{\phi_0}qz, \qquad \ell=\frac{\sqrt{2}\rme^{-\phi_0}}{|q|}.
\end{equation}
Then the periodic identification of the linear example becomes
$t\sim t+2\pi\rmi$, and
\begin{equation}\label{eq:met1}
    \tau_1=\rme^{-\phi_0}(\sigma-\rmi)t, \qquad \rmd s_{E,2}^2=\ell^2\rme^{2u}
\left(\rmd u^2+\rmd v^2\right).
\end{equation}
In the conformal orthonormal frame associated with $(u,v)$, we have
\begin{equation}
    a_{\rm lin}=u+\log\ell, \qquad f_{\rm lin}=-\frac{u}{2}-\frac{\phi_0}{8},
\qquad \epsilon_{\rm lin}=\rme^{f_{\rm lin}}\epsilon_0\,.
\end{equation}

Now apply the harmonic dressing \eqref{eq:harm} with
\begin{equation}
    h=\log\frac{L}{\ell}-u, \qquad \widetilde h=-v, \qquad L>0.
\end{equation}
This affects the conformal frame as follows:
\begin{equation}
    a_{\rm cyl}=\log L, \qquad f_{\rm cyl}=-\frac{u+\rmi v}{2}-\frac{\phi_0}{8}\,.
\end{equation}
Consequently, we get the following metric and spinor
\begin{equation}\label{eq:dressed}
    \rmd s_{E,2}^2=L^2 \left(\rmd u^2+\rmd v^2\right), \qquad\epsilon_{\rm cyl}
=\rme^{-\phi_0/8}\rme^{-t/2}\epsilon_0\,.
\end{equation}
To recover the simple-pole expression \eqref{eq:simple-pole}, introduce a new coordinate $\zeta$ by
\begin{equation}
    \zeta-z_0=L\rme^{-t}\,.
\end{equation}
Then solution \eqref{eq:dressed} is characterized by
\begin{equation}
    \rmd s_{E,2}^2= \frac{L^2}{|\zeta-z_0|^2}\rmd\zeta\,\rmd\bar\zeta\,,
\qquad \tau_1=\rme^{-\phi_0}(\sigma-\rmi) \log\frac{L}{\zeta-z_0}\,.
\end{equation}

On the other hand, all of $\phi,b,U,V,S,T$ are left unchanged. In particular, $F_1, F_5$ and the associated charges are the same as in the linear example, as we also observed in section \ref{sec:constant-scalars}. 

Finally, \eqref{eq:spinor-coordinate-shift} expresses the spinor in the $\zeta$-frame, canceling in particular the phase $-\frac12v$:
\begin{equation}
    \epsilon= \rme^{-\phi_0/8} \left|\frac{\zeta-z_0}{L}\right|^{1/2}\epsilon_0\,,
\end{equation}
as expected for the constant-phase representative of the simple-pole solution. 

The harmonic dressing is essential: the two solutions are not related by a coordinate transformation alone. The linear quotient \eqref{eq:met1} is a punctured flat plane, whereas the dressed metric \eqref{eq:dressed} is a cylinder of constant radius $L$. They have the same underlying topology, but different geometry. Even locally, the metric together with the RR fields distinguishes them, since
\begin{equation}
    \left(g^{ij}\partial_iU\,\partial_jU\right)_{\rm lin} =q^2\rme^{-2u},
\qquad \left(g^{ij}\partial_iU\,\partial_jU\right)_{\rm cyl} =\frac{2\rme^{-2\phi_0}}{L^2}\,.
\end{equation}
Moreover, the linear solution has periodic spin structure, while the cylinder solution has antiperiodic spin structure, which is another indication that they are not related by a coordinate transformation.

\section{More general solutions}\label{sec:general-solutions-appendix}
The local solution space also contains configurations with a non-zero Killing-spinor phase
and solutions in which the dilaton or relative torus size runs. As discussed in section \ref{sec:more-general-solutions},
the natural angular identification leads to a non-trivial monodromy of the Killing spinor,
generically including a monodromy of its norm. Hence the Killing spinor is not globally
single-valued, and we do not consider these configurations as proper global backgrounds.
\subsection{A different kind of cylinder} In comparison to example \eqref{eq:simple-pole}, the simple logarithmic $\tau_1=-\rmi \log z$ leads to a more problematic metric. Setting $\phi_0=b_0=0$ and $\sigma=1$ for simplicity, and using coordinates $z=\rme^{\rho+\rmi\theta}$, we have
\begin{equation}
\tau_1=-\rmi\log z=\theta-\rmi\rho\,, \qquad\tau_2=-\rmi\tau_1=-\rho-\rmi\theta\,.
\end{equation}
Using \eqref{eq:gravi2b} with $\Im f=0$, we find
\begin{equation}
a=\frac{1}{2}\left(\rho+\theta\right)\,,\quad f= -\frac14(\rho +\theta)\,,
\end{equation}
which determines the solution
\begin{equation}\label{eq:generic-log}
    F_1=-\rmi\frac{\rmd z}{z} = -\rmi \rmd \rho + \rmd \theta\,,\qquad
    \rmd s_{E,2}^2=\rme^{\rho+\theta}\rmd z \rmd \bar{z} = \rme^{3\rho+\theta}(\rmd\rho^2 + \rmd \theta^2)\,.
\end{equation}
The field strength $F_1$ is readily interpreted in terms of $Q_{7}$ and $Q_{-1}$ charges \eqref{eq:D-1D7chargeAn}. Yet, the conformal factor is not periodic under $\theta\to\theta+2\pi$. 

This non-periodicity can be removed using the freedom \eqref{eq:harm}. Taking
\begin{equation}
h=\left(\lambda-\frac12\right)\rho-\frac{1}{2}\theta\,,\qquad \widetilde h=\left(\lambda-\frac12\right)\theta+\frac{1}{2}\rho\,,
\end{equation}
the $\theta$-dependence of the metric is canceled for any $\lambda$:
\begin{equation}
    a \to\lambda\rho\,,\qquad f \to -\frac14(\rho +\theta)+\frac{\rmi}{4}\left(\rho+\left(2\lambda-1\right)\theta
\right).
\end{equation}
This results in the following metric and angular dependence of the Killing spinor:
\begin{equation}\label{eq:spinor}
\rmd s_{E,2}^2 =\rme^{2(\lambda+1)\rho} \left(\rmd\rho^2+\rmd\theta^2\right)\,,
\qquad \epsilon_z  =\exp\left( -\frac14(\rho+\theta)  +\frac{\rmi}{4} \left(\rho+(2\lambda-1)\theta\right) \right)\epsilon_0\,.
\end{equation}

For generic $\lambda\neq-1$, the transverse metric is locally flat and
conical. Thus the harmonic dressing replaces the original non-periodic metric by
a flat cone, with opening angle determined by $\lambda$. However, the non-periodicity is still present in the spinor.

It should be noted that the above \eqref{eq:spinor} is still given in the flat frame determined by $z$, so we have not quite performed an ordinary spin frame transformation. Working directly in the cylinder frame determined by $\rho$ and $\theta$ corresponds to
\begin{equation}
    a\to(\lambda+1)\rho\,, \qquad  f\ \to \ -\frac14(\rho +\theta)+ \frac{\rmi}{4} \left(\rho+(2\lambda+1)\theta\right).
    \label{eq:log-cylinder-frame}
\end{equation}
Now $a$ determines the conformal factor of $ \rmd \rho^2 + \rmd\theta^2$ rather than that of $\rmd z \rmd \bar{z}$. The spinor in this case acquires an extra phase
\begin{equation}\label{eq:spinor_cyl}
    \epsilon= \exp\left( -\frac14(\rho+\theta)  +\frac{\rmi}{4} \left(\rho+(2\lambda+1)\theta\right) \right)\epsilon_0\,.
\end{equation}

\paragraph{The cylinder of \cite{Aguilar-Gutierrez:2022kvk}}
The special case $\lambda=-1$ corresponds to having a cylinder. The crucial difference with respect to the previous cylinder \eqref{eq:simple-pole} is the direction of the quotient cycle relative to the RR gradients. The result is the option I solution of \cite{Aguilar-Gutierrez:2022kvk}, which still has a problematic spinor periodicity.\footnote{This issue was not appreciated in \cite{Aguilar-Gutierrez:2022kvk}, because its choice of spinor projectors and chirality was not consistent with the $F_5$ self-duality conventions.}

\subsection{Running solutions with cylindrical asymptotics}\label{sec:running-cylinder}
We now exhibit a deformation of the cylindrical solution discussed in the previous section. This is achieved by making either the dilaton or the relative torus size run according to \eqref{eq:riccati}. 

\paragraph{Running dilaton}
We choose the same holomorphic RR data as in the $\lambda=-1$ logarithmic solution above,
\begin{equation}\label{eq:running-tau}
\tau_1=-\rmi\rme^{-\phi_0}(\rho+\rmi\theta)\,,
\qquad \tau_2=-\rme^{4b_0}(\rho+\rmi\theta)\,,
\end{equation}
but allow the dilaton to run according to \eqref{eq:riccati}. This results in
\begin{equation}\label{eq:running-solution}
\begin{split}
\rme^\phi=\rme^{\phi_0}\tanh\rho\,&,\qquad b=b_0\,,\\
U=\rme^{-\phi_0}\theta\,&,\qquad V=\rme^{-\phi_0}(-\rho+\coth\rho)\,,\\
S=-\rme^{4b_0}\rho\,&,\qquad T=\rme^{4b_0}(1-\theta)\,.
\end{split}
\end{equation}
The transverse metric and Killing spinor are determined via \eqref{eq:gravi2b} and \eqref{eq:gravi1b}. We fix the freedom \eqref{eq:harm} by requiring $\partial_\theta$ to be an isometry and demanding the asymptotic $\theta$-circle to have finite and non-zero circumference, which we will pick to be $2\pi$. This determines the Einstein-frame metric to be
\begin{equation}\label{eq:running-metric}
\begin{split}
\rmd s_E^2={} (1+\rme^{-2\rho}) (\rmd\rho^2+\rmd\theta^2)+\rme^{2b_0}\sum_{i=1}^4\rmd\theta_i^2 +\rme^{-2b_0}\sum_{i=5}^8\rmd\theta_i^2\,,
\end{split}
\end{equation}
while the Killing spinor is given by
\begin{equation}\label{eq:killing_run}
 \epsilon=  (1-\rme^{-4\rho})^{-1/8}  \exp\left( -\frac14(\rho+\theta)
 +\frac{\rmi}{4}(\rho-\theta)  \right)\epsilon_0\,.
\end{equation}
As $\rho\to \infty$, we approach the solution of \cite{Aguilar-Gutierrez:2022kvk} discussed in the previous section. 

The running solution is curved and its transverse Ricci scalar is
\begin{equation}\label{eq:running-curvature}
R_{(2)} =-\frac{4\rme^{-2\rho}}{(1+\rme^{-2\rho})^3}\,,
\end{equation}
which remains finite as $\rho\to 0$. In this limit, the dilaton and RR fields behave as follows:
\begin{equation}
     \rme^\phi\sim\rme^{\phi_0}\rho\,, \qquad \rme^{-\phi}\sim\frac{\rme^{-\phi_0}}{\rho}\,, \qquad V\sim\frac{\rme^{-\phi_0}}{\rho}\,.
\end{equation}
Note that $\rho=0$ lies at finite proper distance.  The running solution describes a flow from a finite-distance weak-coupling endpoint to the cylindrical
D($-$1)/D7 background at $\rho\to\infty$.

For completeness, the same Riccati equation \eqref{eq:riccati} admits the second hyperbolic branch
\begin{equation}
\rme^\phi=\rme^{\phi_0}\coth\rho\,, \qquad V=\rme^{-\phi_0}(-\rho+\tanh\rho)\,,
\end{equation}
with the same $\tau_1,\tau_2$ and the same charges. In this case we get the metric
\begin{equation}
     \rmd s_{E,2}^2  =(1-\rme^{-2\rho})(\rmd\rho^2+\rmd\theta^2)\,,
\end{equation}
which approaches the same cylinder as $\rho\to\infty$, but has a curvature singularity at $\rho=0$. 

There are also reflected solution branches defined on $\rho<0$, approaching the cylinder as $\rho\to-\infty$.

\paragraph{Running torus size}
Finally, let us consider a running torus modulus $b$.  A non-trivial radial torus flow necessarily carries the ``rotated'' D3 charges rather than the D($-$1)/D7 charges of the background considered above. The reason is that since $b=b(H_2)$, a radial flow $b=b(\rho)$ requires $H_2$ to depend on $\rho$. Holomorphicity of $\tau_2=S+\rmi H_2$ then forces $S$ to depend on the conjugate coordinate $\theta$, hence producing D3 monodromy. The solution below therefore approaches the same cylindrical geometry, but not the same RR configuration.

We are therefore led to consider
\begin{equation}\label{eq:running-tau2}
\tau_1=\rme^{-\phi_0}(\rho+\rmi\theta)\,, \qquad \tau_2=-\rmi\rme^{4b_0}(\rho+\rmi\theta)\,.
\end{equation}
This determines
\begin{align}\label{eq:running-solution2}
\rme^\phi=\rme^{\phi_0}\,&,\qquad \rme^{-4b}=\rme^{-4b_0}\tanh\rho\,,\\
U=\rme^{-\phi_0}\rho\,&,\qquad V=\rme^{-\phi_0}(\theta+1)\,, \\
S=\rme^{4b_0}\theta\,&,\qquad T=\rme^{4b_0}(-\rho+\coth\rho)\,.
\end{align}
The resulting Einstein-frame metric is
\begin{equation}\label{eq:running-metric2}
\rmd s_E^2={} \sqrt{1-\rme^{-4\rho}}
(\rmd\rho^2+\rmd\theta^2) +\rme^{2b_0}(\tanh\rho)^{-1/2}
\sum_{i=1}^{4}\rmd\theta_i^2 +\rme^{-2b_0}(\tanh\rho)^{1/2}
\sum_{i=5}^{8}\rmd\theta_i^2\,.
\end{equation}
The gravitino equations \eqref{eq:gravi1b}--\eqref{eq:gravi2b} determine
the full Killing spinor to be
\begin{equation}\label{eq:running-phase2}
\epsilon= (1-\rme^{-4\rho})^{-1/8} \exp\left(\frac14(\theta-\rho)
-\frac{\rmi}{4}(\rho+\theta) \right)\epsilon_0\,.
\end{equation}
For $\rho\to\infty$, we have $b\to b_0$, and the transverse metric approaches the
cylinder \eqref{eq:cylinder}. The interpolation is curved:
\begin{equation}
R_{(2)} =\frac{8\rme^{6\rho}}
{(\rme^{4\rho}-1)^{5/2}}\,,
\end{equation}
and $R_{(2)}\to0$ for $\rho\to\infty$, whereas the endpoint $\rho=0$ is a genuine curvature singularity at finite proper distance. At the same time
$\rme^{4b}\sim\rme^{4b_0}/\rho$, so that one four-torus decompactifies while
the other shrinks.

{\small
\IfFileExists{JHEP.bst}{\bibliographystyle{JHEP}}{\bibliographystyle{unsrt}}
\bibliography{branes_corrected}}

\providecommand{\href}[2]{#2}\begingroup\raggedright\begin{thebibliography}{10}

\bibitem{Aguilar-Gutierrez:2022kvk}
S.~E. Aguilar-Gutierrez, K.~Parmentier and T.~Van~Riet, \emph{{Towards an {\textquotedblleft}AdS$_{1}$/CFT$_{0}${\textquotedblright} correspondence from the D({\ensuremath{-}}1)/D7 system?}}, \href{http://dx.doi.org/10.1007/JHEP09(2022)249}{\emph{JHEP} {\bf 09} (2022) 249}, [\href{http://arxiv.org/abs/2207.13692}{{\tt 2207.13692}}].

\bibitem{Reymond:2024mwe}
S.~Reymond, M.~Trigiante and T.~Van~Riet, \emph{{Supergravity solutions for Dp-D(6 {\ensuremath{-}} p) bound states: from p = 7 to p = {\ensuremath{-}}1}}, \href{http://dx.doi.org/10.1007/JHEP04(2025)037}{\emph{JHEP} {\bf 04} (2025) 037}, [\href{http://arxiv.org/abs/2410.02616}{{\tt 2410.02616}}].

\bibitem{Billo:2021xzh}
M.~Bill{\`o}, M.~Frau, F.~Fucito, L.~Gallot, A.~Lerda and J.~F. Morales, \emph{{On the D({\textendash}1)/D7-brane systems}}, \href{http://dx.doi.org/10.1007/JHEP04(2021)096}{\emph{JHEP} {\bf 04} (2021) 096}, [\href{http://arxiv.org/abs/2101.01732}{{\tt 2101.01732}}].

\bibitem{Nekrasov:2017cih}
N.~Nekrasov, \emph{{Magnificent four}}, \href{http://dx.doi.org/10.4310/ATMP.2020.v24.n5.a4}{\emph{Adv. Theor. Math. Phys.} {\bf 24} (2020) 1171--1202}, [\href{http://arxiv.org/abs/1712.08128}{{\tt 1712.08128}}].

\bibitem{Ibanez:2012zz}
L.~E. Ibanez and A.~M. Uranga, \emph{{String theory and particle physics: An introduction to string phenomenology}}.
\newblock Cambridge University Press, 2, 2012.

\bibitem{Hanany:1996ie}
A.~Hanany and E.~Witten, \emph{{Type IIB superstrings, BPS monopoles, and three-dimensional gauge dynamics}}, \href{http://dx.doi.org/10.1016/S0550-3213(97)00157-0}{\emph{Nucl. Phys. B} {\bf 492} (1997) 152--190}, [\href{http://arxiv.org/abs/hep-th/9611230}{{\tt hep-th/9611230}}].

\bibitem{Danielsson:1997wq}
U.~Danielsson, G.~Ferretti and I.~R. Klebanov, \emph{{Creation of fundamental strings by crossing D-branes}}, \href{http://dx.doi.org/10.1103/PhysRevLett.79.1984}{\emph{Phys. Rev. Lett.} {\bf 79} (1997) 1984--1987}, [\href{http://arxiv.org/abs/hep-th/9705084}{{\tt hep-th/9705084}}].

\bibitem{Witten:2000mf}
E.~Witten, \emph{{BPS Bound states of D0 - D6 and D0 - D8 systems in a B field}}, \href{http://dx.doi.org/10.1088/1126-6708/2002/04/012}{\emph{JHEP} {\bf 04} (2002) 012}, [\href{http://arxiv.org/abs/hep-th/0012054}{{\tt hep-th/0012054}}].

\bibitem{Janssen:1999sa}
B.~Janssen, P.~Meessen and T.~Ortin, \emph{{The D8-brane tied up: String and brane solutions in massive type IIA supergravity}}, \href{http://dx.doi.org/10.1016/S0370-2693(99)00315-9}{\emph{Phys. Lett. B} {\bf 453} (1999) 229--236}, [\href{http://arxiv.org/abs/hep-th/9901078}{{\tt hep-th/9901078}}].

\bibitem{Massar:1999sb}
M.~Massar and J.~Troost, \emph{{D0 - D8 - F1 in massive IIA SUGRA}}, \href{http://dx.doi.org/10.1016/S0370-2693(99)00660-7}{\emph{Phys. Lett. B} {\bf 458} (1999) 283--287}, [\href{http://arxiv.org/abs/hep-th/9901136}{{\tt hep-th/9901136}}].

\bibitem{Imamura:2001cr}
Y.~Imamura, \emph{{1/4 BPS solutions in massive IIA supergravity}}, \href{http://dx.doi.org/10.1143/PTP.106.653}{\emph{Prog. Theor. Phys.} {\bf 106} (2001) 653--670}, [\href{http://arxiv.org/abs/hep-th/0105263}{{\tt hep-th/0105263}}].

\bibitem{Bergshoeff:2003sy}
E.~Bergshoeff, U.~Gran, R.~Linares, M.~Nielsen and D.~Roest, \emph{{Domain walls and the creation of strings}}, \href{http://dx.doi.org/10.1088/0264-9381/20/15/313}{\emph{Class. Quant. Grav.} {\bf 20} (2003) 3465--3482}, [\href{http://arxiv.org/abs/hep-th/0303253}{{\tt hep-th/0303253}}].

\bibitem{Brezin_ZJ_1992}
E.~Brézin and J.~Zinn-Justin, \emph{Renormalization group approach to matrix models}, \href{http://dx.doi.org/10.1016/0370-2693(92)91953-7}{\emph{Physics Letters B} {\bf 288} (Aug., 1992) 54–58}.

\bibitem{Zinn-Justin:2014wva}
J.~Zinn-Justin, \emph{{Random vector and matrix and vector theories: a renormalization group approach}}, \href{http://dx.doi.org/10.1007/s10955-014-1103-y}{\emph{J. Statist. Phys.} {\bf 157} (2014) 990--1016}, [\href{http://arxiv.org/abs/1410.1635}{{\tt 1410.1635}}].

\bibitem{Guan:2026sjw}
X.~Guan, J.~Karlsson, S.~Reymond and T.~Van~Riet, \emph{{The IKKT renormalization group flow is IIB: toward zero-d holography}},  \href{http://arxiv.org/abs/2606.27333}{{\tt 2606.27333}}.

\bibitem{Bergshoeff:2000qu}
E.~Bergshoeff and A.~Van~Proeyen, \emph{{The Many faces of OSp(1|32)}}, \href{http://dx.doi.org/10.1088/0264-9381/17/16/312}{\emph{Class. Quant. Grav.} {\bf 17} (2000) 3277--3304}, [\href{http://arxiv.org/abs/hep-th/0003261}{{\tt hep-th/0003261}}].

\bibitem{Burgess:1989da}
C.~P. Burgess and A.~Kshirsagar, \emph{{Wormholes and Duality}}, \href{http://dx.doi.org/10.1016/0550-3213(89)90186-7}{\emph{Nucl. Phys. B} {\bf 324} (1989) 157--166}.

\bibitem{Gibbons:1995vg}
G.~W. Gibbons, M.~B. Green and M.~J. Perry, \emph{{Instantons and seven-branes in type IIB superstring theory}}, \href{http://dx.doi.org/10.1016/0370-2693(95)01565-5}{\emph{Phys. Lett. B} {\bf 370} (1996) 37--44}, [\href{http://arxiv.org/abs/hep-th/9511080}{{\tt hep-th/9511080}}].

\bibitem{Loges:2022nuw}
G.~J. Loges, G.~Shiu and N.~Sudhir, \emph{{Complex saddles and Euclidean wormholes in the Lorentzian path integral}}, \href{http://dx.doi.org/10.1007/JHEP08(2022)064}{\emph{JHEP} {\bf 08} (2022) 064}, [\href{http://arxiv.org/abs/2203.01956}{{\tt 2203.01956}}].

\bibitem{Hertog:2024nys}
T.~Hertog, S.~Maenaut, B.~Missoni, R.~Tielemans and T.~Van~Riet, \emph{{Stability of axion-saxion wormholes}}, \href{http://dx.doi.org/10.1007/JHEP11(2024)151}{\emph{JHEP} {\bf 11} (2024) 151}, [\href{http://arxiv.org/abs/2405.02072}{{\tt 2405.02072}}].

\bibitem{lander}
L.~Umans, \emph{Supergravity solution for the {D}$(-1)$/{D7} bound state},  Master's thesis, KU Leuven, 2025.

\bibitem{Greene:1989ya}
B.~R. Greene, A.~D. Shapere, C.~Vafa and S.-T. Yau, \emph{{Stringy Cosmic Strings and Noncompact Calabi-Yau Manifolds}}, \href{http://dx.doi.org/10.1016/0550-3213(90)90248-C}{\emph{Nucl. Phys. B} {\bf 337} (1990) 1--36}.

\bibitem{Horowitz:1991cd}
G.~T. Horowitz and A.~Strominger, \emph{{Black strings and P-branes}}, \href{http://dx.doi.org/10.1016/0550-3213(91)90440-9}{\emph{Nucl. Phys. B} {\bf 360} (1991) 197--209}.

\bibitem{Tseytlin:1996bh}
A.~A. Tseytlin, \emph{{Harmonic superpositions of M-branes}}, \href{http://dx.doi.org/10.1201/9781482268737-28}{\emph{Nucl. Phys. B} {\bf 475} (1996) 149--163}, [\href{http://arxiv.org/abs/hep-th/9604035}{{\tt hep-th/9604035}}].

\bibitem{Tseytlin:1997cs}
A.~A. Tseytlin, \emph{{Composite BPS configurations of p-branes in ten-dimensions and eleven-dimensions}}, \href{http://dx.doi.org/10.1088/0264-9381/14/8/009}{\emph{Class. Quant. Grav.} {\bf 14} (1997) 2085--2105}, [\href{http://arxiv.org/abs/hep-th/9702163}{{\tt hep-th/9702163}}].

\bibitem{Aharony:2024fid}
O.~Aharony and E.~Y. Urbach, \emph{{Type II string theory on AdS3{\texttimes}S3{\texttimes}T4 and symmetric orbifolds}}, \href{http://dx.doi.org/10.1103/PhysRevD.110.046028}{\emph{Phys. Rev. D} {\bf 110} (2024) 046028}, [\href{http://arxiv.org/abs/2406.14605}{{\tt 2406.14605}}].

\bibitem{Gautason:2025plx}
F.~F. Gautason and J.~van Muiden, \emph{{Ensembles in M-theory and holography}}, \href{http://dx.doi.org/10.1007/JHEP11(2025)078}{\emph{JHEP} {\bf 11} (2025) 078}, [\href{http://arxiv.org/abs/2505.21633}{{\tt 2505.21633}}].

\bibitem{Becker:2006dvp}
K.~Becker, M.~Becker and J.~H. Schwarz, \emph{{String theory and M-theory: A modern introduction}}.
\newblock Cambridge University Press, 12, 2006, \href{http://dx.doi.org/10.1017/CBO9780511816086}{10.1017/CBO9780511816086}.

\end{thebibliography}\endgroup

\end{document}